\documentstyle[11pt]{article}

         \def\be{\begin{equation}}
         \def\ba{\begin{array}}
         \def\bea{\begin{eqnarray}}
         
         \def\ee{\end{equation}}
         \def\ea{\end{array}}
         \def\eea{\end{eqnarray}}
         \def\Rset{\rm {I\kern-.200em R}}
         \def\Cset{\rm {I\kern-.520em C}}
         \def\R#1{\Rset^{#1}}
         
         \def\GL#1{{\rm GL}( #1 ) }
         
         \def\gl#1{{\rm gl}( #1 ) }
         \def\sl#1{{\rm sl}( #1 ) }
         \def\half{{1 \over 2}}
         \def\GLq#1{{\rm GL}_q( #1 ) }
         \def\IGL#1{{\rm IGL}( #1 ) }
         \def\igl#1{{\rm igl}( #1 ) }
         \def\IGLq#1{{\rm IGL}_q( #1 ) }
         \def\Uq#1{{\rm U}_q \left( #1 \right) }
         \def\U#1{{\rm U} \left( #1 \right) }

         \def\Xs#1#2{{\rm X}^{#1}_{#2} }
         \def\Xc#1#2#3{{\rm X}^{#1}_{#2 \, #3} }

         \def\H#1{ {\rm H}_{{#1}} }
         \def\E#1#2{{\rm e}_{#1 \, #2} }
         \def\Rq#1{\Rset_q^{#1}}
         \def\p#1#2{p_{#1 \, #2}}
\begin{document}
\rightline{IPM-94-47}
\rightline{hep-th/9408059}
\vskip 10mm  \noindent
{\large
 {\bf
   On The Inhomogeneous Quantum Groups
      } 
 } 
\vskip 10 mm  \noindent
             A. Shariati ${}^{\dagger,\ddagger}$,
             A. Aghamohammadi ${}^{\diamond,\ddagger}$
\vskip 10 mm
{\it
\noindent
             $^\dagger$
             Department of Physics, Sharif University of Technology,
             P.O.Box 9161, Tehran 11365, Iran.
\\ \noindent
             $^\ddagger$
             Institute for Studies in Theoretical Physics and
              Mathematics,
             P.O.Box  5746, Tehran 19395, Iran.
\\ \noindent
              $^\diamond$
              Department of Physics, Alzahra University
              P.O.Box 19834, Tehran, Iran
    } 
\vskip 10 mm

\begin{abstract}
Using the multi-parametric deformation of the algebra of functions on
$ \GL{n+1} $ and the universal enveloping algebra $ \U{\igl{n+1}} $,
we construct the multi-parametric quantum groups $ \IGLq{n} $ and
$ \Uq{\igl{n}} $.
\end{abstract}
\vskip 15 mm

   The quantization of inhomogeneous groups is not well stablished.
These non-semisimple groups are important, both from a purely
mathematical point of view , and also because one of the most important groups
appearing in physics, viz. the Poincar\'e group, is inhomogeneous.
Various authors in the last few years studied the quantization of these
objects from different viewpoints and by different methods
[1-7].

   In this paper we consider the inhomogeneous general linear group
$\IGL{n}$. We will show that in analogy with the classical case,
the multi-parameteric deformation of the parent $\GL{n+1}$ will lead
to a consistent quantization of both the algebra of functions and the
universal enveloping algebra of the inhomogeneous group $ \IGL{n} $.
Multi-parametric quantum groups
have been studied by various authors [8-14]. In this paper we
will use the
notations of Ref. [13].
To clarify our approach, we will first consider the two dimensional case
$\IGL{2}$;
then we will generalize to the higher dimensions.
Throughout this paper, when we use a
subscript $q$ for an object, we mean the multi-parameteric deformation
of that object. The number of, and the conditions on, the parameters will be
clear from the context.

   Let us consider the two dimensional inhomogeneous group $\IGL{2}$.
Its action on the two dimensional plane $\R{2}$ is:
\be
    x \longmapsto x' = a x + b y + u
    \qquad
    y \longmapsto y' = c x + d y + v .
\ee
This action may be represented on the $ z = 1 $
subset of $ \R{3} $ by means of the following matrix:
\be
  \pmatrix{ x \cr y \cr 1 }
  \longmapsto
  \pmatrix{ x' \cr y' \cr 1}
  =
  \pmatrix{ a & b & u \cr
            c & d & v \cr
            0 & 0 & 1
           }
  \pmatrix{ x \cr y \cr 1}.
\ee
{}From this representation we can easily write the Lie algebra spanned by:
\be
 \ba{lll}
  H_{1} = \pmatrix{
                    1 &  0  &  0  \cr
                    0 & -1  &  0  \cr
                    0 &  0  &  0
                    }
\quad &
  X_{1}^{+} = \pmatrix{
                         0  &  1  &  0  \cr
                         0  &  0  &  0  \cr
                         0  &  0  &  0
                      }
\quad  &
  X_{1}^{-} = \pmatrix{
                        0  &  0  &  0  \cr
                        1  &  0  &  0  \cr
                        0  &  0  &  0
                        }
  \cr \\
  X_{2}^{+} = \pmatrix{
                        0  &  0  &  0  \cr
                        0  &  0  &  1  \cr
                        0  &  0  &  0
                        }
\quad  &
  X_{12}^{+} = \pmatrix{
                         0  &  0  &  1  \cr
                         0  &  0  &  0  \cr
                         0  &  0  &  0
                         }
\quad  &
  {\rm K} = \pmatrix{
                1  &  0  &  0  \cr
                0  &  1  &  0  \cr
                0  &  0  &  0
                 }
{}.
 \ea
\ee
The set $ \{ H_{1}, X_{1}^{+} , X_{1}^{-} \} $ generates $ SL(2) $;
$ {\rm K} $ generates dilatations (scaling), and, $ X_{2}^{+} $ and $
X_{12}^{+} $
are the generators of the translations.
Note that this Lie algebra, $\igl{2}$, is a subalgebra of the
$\gl{3}$, with the other generators of $\gl{3}$ being,
\bea
  \label{3}
  H_{2} = \pmatrix{
                   0 & 0 & 0 \cr
                   0 & 1 & 0 \cr
                   0 & 0 & -1
                   }
 \quad &
  X_{2}^{-} = \pmatrix{
                        0  &  0  &  0  \cr
                        0  &  0  &  0  \cr
                        0  &  1  &  0
                        }
\quad  &
  X_{12}^{-} = \pmatrix{
                         0  &  0  &  0  \cr
                         0  &  0  &  0  \cr
                         1  &  0  &  0
                         }
{}.
\eea
The identity matrix $I$, which is a generator of $\gl{3}$, is not an
independent
operator and we have
\be
  \label{5}
 {\rm I}={3\over 2} {\rm K} - {1 \over 2} H_{1} - H_{2} .
\ee
We would like to study the quantum deformed version of these relations.
First of all, we have to consider a two dimensional quantum space $\Rq{2}$
spanned by $(x,y)$. If this non-commutative space is to be embedded in a three
dimensional quantum space $\Rq{3}$ with coordinates $(x,y,z)$
by setting $z=1$
,then, $z$ must commute with everything, otherwise it would be meaningless to
set $z=1$. This is only possible if we use the multi-parameteric quantum space
$\Rq{3}$ with a special choice of the parameters.
To see this we write the
commutation relations of the coordinates of $\Rq{3}$:
\be
  x_{i} \, x_{j} = q \, p_{ij} \, x_{j} \, x_{i} \qquad i < j
\ee
Therefore, $ x_{3} = z $ is central only if $ p_{13} = p_{23} = q^{-1} $.

  The multi-parameteric quantum group $\GLq{3}$ [13] is generated by the
entries of the following $T$ matrix.
\be
  T = \pmatrix{
    a & b & u \cr
    c & d & v \cr
    e & f & w \cr
                }
{}.
\ee
If we set $e=f=0$, then $w$ will automatically commute with $a,b,c$ and $d$.
However, $w$ has non-trivial commutation relations with $u$ and $v$. To make
$w$ central, we have to impose the same condition
$ \p{1}{3}=\p{2}{3}=q^{-1} $.
 we call the resulting quantum group, $ {\rm IGL}_{q}(2) $.

For completeness we write the commutation relations among the generators,
\be
 \ba{lll}
  a \, b = q p^{-1} \, b \, a                           &
\quad  a \, u = q^{2} \, u \, a                              &
\quad  b \, u = q^{2} \, u \, b                              \cr
  c \, d = q \, p^{-1} \, d \, c                        &
\quad  c \, v = q^{2} \, v \, c                              &
\quad  d \, v = q^{2} v \, d                                 \cr
  a \, c = q \, p \, c \, a                             &
\quad  b \, d = q \, p \, d \, b                             &
\quad  u \, v = q \, p \, v \, u                             \cr
  b \, c = p^{2} \, c \, b                              &
\quad  u \, c = q^{-1} \, p \, c \, u                        &
\quad  u \, d = q^{-1} \, p \, d \, u                        \cr
  a \, d - d \, a = \lambda \, p^{-1} \, b \, c         &
\quad  a \, v - q \, p \, v \, a = q \, \lambda \, u \, c    &
\quad  b \, v - q \, p \, v \, b = q \, \lambda \, u \, d    \cr
  [ w , \cdots ] = 0                                   \cr
 \ea
\ee
where $ p := \p{1}{2} $.
The co-product, co-unity and the antipode for this quantum group are also
inherited from ${\rm GL}_{q}(3)$.

  Generalizing this to $n$ dimensions is straightforward. Use the
multi-parameteric quantum space $\Rq{n+1}$ and the corresponding quantum
group $ \GLq{n+1} $, then set $ \p{i}{n+1}=q^{-1} $ for $ i=1,2,\dots,n $;
we set $T_{{n+1} \, i} = 0$ for $i=1,2,\dots,n$ and then because of the
choice of parameters, $x_{n+1} \, \in \,\Rq{n+1}$ and $w:=T_{{n+1} \,
{n+1}} \, \in \, \GLq{n+1}$ will become central. Setting $x_{n+1} = 1$ and $w
= 1$, one gets the quantum group $\IGLq{n}$.

  Now we turn to the quantization of the universal enveloping algebra
$\U{\igl{2}}$. There exists a multi-parameteric deformation of $\U{\gl{3}}$
which we name $\Uq{\gl{3}}$. This Hopf algebra is generated by 9 generators
$\H{1},\H{2},\Xs{\pm}{1},\Xs{\pm}{2},\Xc{\pm}{1}{2}$ and $I$. Following
the analogy with the classical case, we expect that in this
quantum case too, the subalgebra generated by
$ \{ \H{1}, {\rm K}, \Xs{\pm}{1}, \Xs{+}{2}, \Xc{+}{1}{2} \} $
is the quantum group $\Uq{\igl{2}}$.
Let's examine this idea.

  The commutation relations: $\Xs{-}{2}$ and
$\Xc{-}{1}{2}$ do not appear in the commutation relations of the set
$ \{ \H{1}, {\rm K}, \Xs{\pm}{1}, \Xs{+}{2}, \Xc{+}{1}{2} \} $ where  ${\rm K}$
is given
by (\ref{5}).
\be
 \ba{ll}
  [ \H{1} , {\rm K} ] = 0                                       &
\quad  [ {\rm K}, \Xs{\pm}{1} ] = 0                                   \cr
  [ {\rm K} , \Xs{+}{2} ] = 0                                    &
\quad   [ \Xs{+}{1}, \Xc{+}{1}{2} ]_{q} = 0                            \cr
  [ \Xs{+}{2} , \Xc{+}{1}{2} ]_{q^{-1}} = 0                      &
\quad   [ \H{1} , \Xs{\pm}{1} ] = \pm 2 \Xs{\pm}{1}                   \cr
  [ \H{1} , \Xc{+}{1}{2} ] =  \Xc{\pm}{1}{2}                    &
\quad   [ {\rm K} , \Xc{+}{1}{2} ] =  \Xc{\pm}{1}{2}                   \cr
  [ \Xs{+}{1} , \Xs{-}{1} ] = [ \H{1} ]_{q}                     &
\quad   [ \Xs{-}{1} , \Xc{+}{1}{2} ] = q^{\half + \H{1}} \Xs{+}{2}.   \cr
 \ea
\ee
Here we have used the notations:
\be
  [ x ]_{q} := { { q^{x} - q^{-x} } \over { q - q^{-1} } }
\ee
\be
  [ x , y ]_{q} := q^{\half} x \, y - q^{-\half} y \, x .
\ee

So as an algebra $\Uq{\igl{2}}$ is a sub-algebra of $\Uq{\gl{3}}$
but as Hopf algebra, it isn't a sub-Hopf-algebra of $\Uq{\igl{3}}$.
Let's consider the co-product of simple roots of the multi-parametric
$\Uq{\gl{n+1}}$,
[13].
\be
  \Delta(\Xs{\pm}{i}) =
  \Xs{\pm}{i} \otimes q^{\half \H{i}} \big( \, \prod_{j=1}^{n} s_{ij}^{\pm
  \sum_{k=1}^{n} a_{jk}^{-1} \H{k} }\big) \,  s_{i}^{\pm {\rm I}}
  +
   q^{-\half \H{i}} \big( \, \prod_{j=1}^{n} s_{ij}^{\mp
  \sum_{k=1}^{n} a_{jk}^{-1} \H{k} }\big) \,  s_{i}^{\mp {\rm I}} \otimes
\Xs{\pm}{i} \ee
where $ [a_{ij}] $ is the Cartan matrix of $\sl{n+1}$, and $a^{-1}_{ij}$
stands for $ij$th element of inverse of cartan matrix and their explicit form
are as below
\be
a_{ij}=2\delta_{ij}\,-\, \delta_{i+1,j}\, -\, \delta_{i-1,j}
\ee
\be  \label {55}
a^{-1}_{ij}={j(n+1-i)\over n+1}\, - \, (j-1)\theta _{ij},\quad \theta _{ij}=
\{^{1\qquad i<j}_{0\qquad i\geq j}
\ee
$s_{ij}$  and $s_{i}$ are extra parameters of deformation. $ s_{ij} =
{s_{ji}}^{-1} $
and $s_{ii} = 1$. Therefore, the number of
parameters is $ 1 + \half n (n+1) $.
As we see the generators $\H{2}$ and $\rm I$
appear in the co-product of $X_1^{\pm }$. If we fix the parameters in a special
way, then extra generators in the co-product disappear. We can construct
$\Uq{\igl{2}}$ as a sub-Hopf algebra of  $\Uq{\gl{3}}$ with the following
choice for parameters.
\be
s_{12}:=s,\quad s_1=s^{2\over 3},\quad s_2=q^{1\over 2}s^{-1\over 3}
\ee
Then the co-products of the generators of $\Uq{\igl{2}}$ are
\be
 \ba{l}
  \Delta(\Xs{\pm}{1}) = \Xs{\pm}{1} \otimes q^{\half \H{1} } \, s^{ \pm \half
(\H{1}-  {\rm K}) } + q^{ - \half \H{1} } \, s^{ \mp \half (\H{1}-{\rm K})
}\otimes \Xs{\pm}{1}   \cr
  \Delta( \Xs{+}{2} ) = \Xs{+}{2} \otimes (  q^{ \half} \, s )^{-\half \H{1}}
  \, ( q^{ 3 \over 2 } \, s^{-1} )^{\half {\rm K}}
  + (  q^{ \half} \, s )^{\half \H{1}} \, ( q^{ 3 \over 2 } \, s^{-1} )^{-\half
{\rm K}}   \otimes \Xs{+}{2}
  \cr
  \Delta( \H{1} ) = \H{1} \otimes 1 + 1 \otimes \H{1}
  \cr
  \Delta( {\rm K} ) = {\rm K} \otimes 1 + 1 \otimes {\rm K} .
 \ea
\ee
The co-product of $\Xc{+}{1}{2}$ may be obtained from these relations,
commutation relations, and the fact that $\Delta$ is a homomorphism of algebra.

   The antipode: This is also inherited from the $\Uq{\gl{3}}$ and our choice
of parameters.
\be
 \ba{lll}
  S ( \H{1} )       = - \H{1}                   &
\quad  S( {\rm K} )            = - {\rm K}             &
\quad   S( \Xs{-}{1} )    = - q^{-2} \, \Xs{-}{1}     \cr
  S( \Xs{+}{1} )    = - q^{2}  \, \Xs{+}{1}     &
\quad   S( \Xs{+}{2} )    = - q^{2}  \, \Xs{+}{2}     &
\quad   S( \Xc{+}{1}{2} ) = - q^{4}  \, \Xc{+}{1}{2}. \cr
 \ea
\ee
At this point it is convenient to introduce another notation for the
generators which reflects their ``geometric'' meaning better,
\be
  ( {\rm K , J_3   , J_+       , J_-       , P_1          , P_2 }      )
  :=
  ( {\rm K} , \H{1} , \Xs{+}{1} , \Xs{-}{1} , \Xc{+}{1}{2} , \Xs{+}{2} ).
\ee
  Now we generalize this method to higher dimensions.
The procedure is as follows.
   We use the multi-parameteric deformation
of $\U{\gl{n+1}}$ which depends on $ 1 + \half n(n+1) $ parameters
\be
  \{ q , s_{ij} , i , j = 1 , 2 , \dots , n \}
\ee
and $ n^2 $ generators
\be
  \{ \H{i} ,{\rm I}, \Xc{\pm}{i}{j} , i \leq j = 1 , 2 , \dots , n \}.
\ee
Here, for notational convenience we have named $\Xc{\pm}{i}{i} := \Xs{\pm}{i}$,
which are simple roots, and $\Xc{\pm}{i}{j}$ for $i\ne j $ are
composite roots.
The matrices of the generators, in the un-deformed case are:
\be
  \Xc{+}{i}{j} = \E{i}{j+1}, \quad
  \Xc{-}{i}{j} = \E{j+1}{i}, \quad
  \H{i} = \E{i}{i} - \E{i+1}{i+1}, \quad
 {\rm I}= \sum_{i=1}^{n+1} \E{i}{i} .
\ee
In the above expressions, $\E{i}{j}$ are the standard basis of matrices,
with entries:
\be
  ( \E{i}{j} )_{m \, n} = \delta_{i \, m} \, \delta_{j \, n} .
\ee
$\Uq{\gl{n}}$ is generated by the following subset of generators:
\be
  \label{22}
\{H_i , {\rm K} , X_{ij}^{\pm},\qquad i\leq j=1,2,\cdots ,n-1\}
\ee
where
\be
{\rm K}={1\over n+1}\sum_{i=1}^n{i \H{i} } + {n \over n+1}{\rm I}.
\ee

A basis for $\igl{n}$ is obtained by adding to the set (\ref{22}) the following
translators:
\be
{\rm P_i}:=\Xc{+}{i}{n} \qquad i=1,2,\cdots ,n .
\ee
In the quantum case also,  we introduce this set as the generating set
of the multi-parametric $\Uq{\igl{n}}$.
\be
{\cal B }=\{{\rm K},\, \H{i},\, \Xc{\pm}{i}{j},\,
\quad i\leq j=1,2,\cdots ,n-1,\,  {\rm P_k},\quad k\leq \,n \,  \}
\ee
${\cal B }$ is a subset of the multi-parametric $\Uq{\gl{n+1}}$. It
is straightforward to check that this set is closed under ``commutation''.

Looking at the co-product one sees that $\H{n}$ and ${\rm I}$ appear in some
of the terms,
e.g. $\Delta(\Xs{\pm}{i}) $. However, we have the freedom of fixing the
parameres $s_{ij}$ and $s_i$ in such a way that only $\H{j}$ for $j<n+1$ and
${\rm K}$ appear in these expressions.
If we fix the prameters $s_{ij}$ and $s_i$ as below
\be
  s_{i} =q^{\half \delta_{in}} \prod_{j=1}^n s_{ij}^{a_{jn}^{-1}}
\ee
the generators $ \H{n}$ and $\rm I$ disappear from the
co-products.
This will fix $n$ parameters and we are left with $1+ \half n(n-1)$ free
parameters.
We have checked that co-unity and antipode also respect this splitting
of generators
of $\Uq{\gl{n+1}}$ for the above mentioned choice of parameters.

In summary, using the multi-parametric deformation of $\U{\gl{n+1}}$, a special
choice of the parameters lead to a Hopf algebra that contains a
sub-Hopf-algebra
generated by ${\cal B}$, which we call it $\Uq{\gl{n}}$. It depends on
$1+ \half n(n-1)$ parameters.
 $\Uq{\igl{n}}$ is generated by the set of generators of its
subalgebra $\Uq{\gl{n-1}}$ and $n$ translations $ P_i := \Xc{+}{i}{n} $.
The commutation relations
are the following, note that we have written the sub-algebra $\gl{n}$ in the
Chevalley basis
 \be
  \ba{ll}
    [\H{i} ,\H{j} ]=0   &
   \qquad [ {\rm K},\cdots ]=0 \cr
    [\H{i},X_j^{\pm}]=\pm a_{ij} X_j^{\pm} &
   \qquad    [X_i^+,X_j^-]=\delta_{ij} [\H{i}]
  \ea
\ee
\be
    (X_{i\pm 1}^{\pm})^2X_i^{\pm} -(q+q^{-1})X_{i\pm 1}^{\pm}X_i^{\pm}X_{i\pm
      1}^{\pm}+X_i^{\pm} (X_{i\pm 1}^{\pm})^2=0
\ee
\be
  \ba{ll}
    [\H{i},P_j]=\sum_{k=j}^n a_{ik} P_j &
   \qquad    [{\rm K},P_i]=P_i \cr
    [X_i^+,P_i]_q=0 &
   \qquad    [P_i, X_j^-]=\delta _{ij}q^{-\half}P_{i+1}q^{\H{i}} \cr
    [P_i,X_{i-1}^+]_q=P_{i-1} &
   \qquad    [P_i,X_j^+]=0, \qquad \vert i-j\vert \leq 2
  \ea
\ee
The co-products are
\be
  \Delta( \H{i} ) = \H{i} \otimes 1 + 1 \otimes \H{i}
\ee
\be
  \Delta( {\rm K} ) = {\rm K} \otimes 1 + 1 \otimes {\rm K} .
\ee
\be  \ba{ll}
   \Delta(\Xs{\pm}{i})\, =\, &\Xs{\pm}{i}\otimes q^{\half
\H{i}}\big(\, \prod_{j=1}^{n}
s_{ij}^{\pm \big[ \, \sum_{k=1}^{n-1}({k(n-1)\over n})-(k-1)\theta_{jk})\H{k} +
(1-j+{1\over n}){\rm K}\, \big] }\, \big) \, +
\cr & \cr & q^{-\half \H{i}}\big( \, \prod_{j=1}^{n}
s_{ij}^{\mp \big[ \, \sum_{k=1}^{n-1}({k(n-1)\over n})-(k-1)\theta_{jk})\H{k} +
(1-j+{1\over n}){\rm K}\, \big]}\, \big)
\otimes \Xs{\pm}{i}
\ea  \ee

\be  \ba{ll}
\Delta({\rm P_n})=&{\rm P_n}\otimes q^{-\half
\sum_{k=1}^{n-1}{k\over n} \H{n} }\big( \, \prod_{j=1}^{n}
s_{nj}^{ \sum_{k=1}^{n-1}({k(n-1)\over n})-(k-1)\theta_{jk})}\, \big) ^{\H{k} }
\big( \, q^{{n+1\over 2n}}(\prod_{j=1}^n s_{nj}^{1-j+{1\over n}})\,\big) ^{\rm
K}+\cr & \cr  & q^{\half \sum_{k=1}^{n-1}{k\over n} \H{n} }\big(
\, \prod_{j=1}^{n}
s_{nj}^{ \sum_{k=1}^{n-1}({-k(n-1)\over n}+(k-1)\theta_{jk})}\big) ^{\H{k} }
\big( \, q^{{-n+1\over 2n}}(\prod_{j=1}^n s_{nj}^{-1+j-{1\over n}})\big) ^{\rm
K} \otimes {\rm P_n}
 \ea \ee
for computing these co-products one must use (\ref {55}).
The co-products of the other ${\rm P_i}$s can be obtained by using the fact
that $\Delta $  is a homomorphism of the algebra and using the following
relation
\be
{\rm P_i}=[\cdots [{\rm
P_n},\Xs{+}{n-1}]_q,\Xs{+}{n-2}]_q,\Xs{+}{n-3}]_q,\cdots \Xs{+}{i}]_q .
\ee

After the submission of this paper we have been noted that multiparametric
deformation of the algebra of functions $ \IGLq{n} $ and its
differential calculus have  been studied in \cite{Cas 2} where the authors
found constraints on the parameters by taking dilatations in the center of the
algebra.

 
\end{document}